\documentclass[conference]{IEEEtran}
\IEEEoverridecommandlockouts

\usepackage{cite}
\usepackage{amsmath,amssymb,amsfonts}
\usepackage{algorithmic}
\usepackage{graphicx}
\usepackage{textcomp}
\usepackage{color}
\usepackage[ruled,vlined,shortend,linesnumbered]{algorithm2e}
\usepackage[numbers,sort&compress]{natbib}
\def\BibTeX{{\rm B\kern-.05em{\sc i\kern-.025em b}\kern-.08em
    T\kern-.1667em\lower.7ex\hbox{E}\kern-.125emX}}
    
\begin{document}

\title{FactorHD: A Hyperdimensional Computing Model for Multi-Object Multi-Class Representation and Factorization\\
}

 \author{
 \small
 Yifei Zhou$^{1}$, Xuchu Huang$^{1}$, Chenyu Ni$^1$, Min Zhou$^{1}$, Zheyu Yan$^{1,2,*}$, Xunzhao Yin$^{1,2,*}$, Cheng Zhuo$^{1,2,*}$\\
 $^1$Zhejiang University, Hangzhou, China;
 $^2$Key Laboratory of Collaborative Sensing and
Autonomous Unmanned Systems of \\ Zhejiang Province, China;
$^*$Corresponding authors, email: \{zyan2, czhuo, xzyin1\}@zju.edu.cn
\vspace{-2ex}
}

\maketitle

\begin{abstract}
Neuro-symbolic artificial intelligence (neuro-symbolic AI) excels in logical analysis and reasoning. 
Hyperdimensional Computing (HDC), a promising brain-inspired computational model, is integral to neuro-symbolic AI. 
Various HDC models have been proposed to represent class-instance and class-class relations, but when representing the more complex class-subclass relation, where multiple objects associate  different levels of classes and subclasses, they face challenges for factorization,
a crucial task for neuro-symbolic AI systems.
In this article, we propose FactorHD, a novel HDC model capable of representing and factorizing the complex class-subclass relation efficiently.
FactorHD features a symbolic encoding method that embeds an extra memorization clause, preserving more information for multiple objects. In addition, it employs an efficient factorization algorithm that selectively eliminates  redundant classes by  identifying the  memorization clause of the target class.
Such model significantly enhances computing efficiency and accuracy in representing and factorizing multiple objects with class-subclass relation,
overcoming limitations of existing HDC models such as "superposition catastrophe" and "the problem of 2".
Evaluations show that FactorHD achieves approximately $5667\times$ speedup at a representation size of $10^9$ compared to existing HDC models.
When integrated with the ResNet-18 neural network, FactorHD achieves $92.48\%$ factorization accuracy on the Cifar-10 dataset.
\end{abstract}


\vspace{-1ex}
\section{Introduction}
\label{sec:introduction}
\begin{figure*}
     \centering
     \vspace{-0.5em}
     \includegraphics[width=1\linewidth]{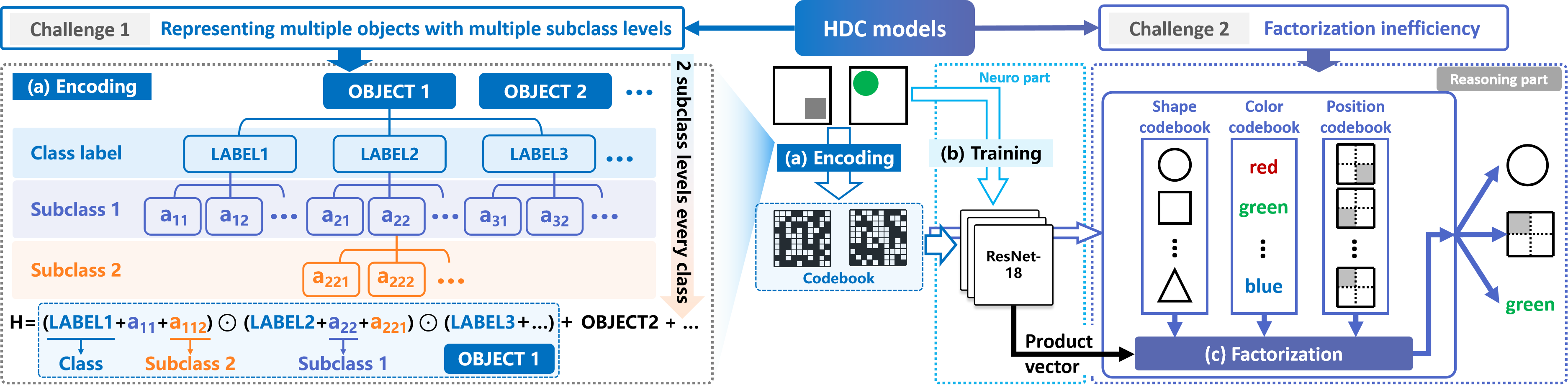}
     \vspace{-2em}
     \caption{
     (a) An example of class-subclass relation with 3 classes (labels) and 2 subclass levels in hierarchy. FactorHD represents this relation by bundling subclass levels within the same class and binding different classes together.
     The encoding process generates HV codebooks.
     Existing HDC models struggle with such complex representations. (b) The training processes involve the implementation of ResNet-18 network as the neuro part of neuro-symbolic AI model. (c) The factorization process given an object query. The HDC challenges involve representation and factorization.}
 \label{fig:challenge}
 \vspace{-1.5em}
 \end{figure*}

Neuro-symbolic artificial intelligence (neuro-symbolic AI) combines symbolic methods with modeling techniques to analyze and reason about problems logically \cite{Chaudhuri2021}. 
Unlike traditional neural structures that rely on repeated training and weighted connections between neurons, neuro-symbolic AI employs explicit representations that are comprehensible to human brains \cite{Sarker2022}, 
leverages hyperdimensional computing (HDC) as the computational paradigm,
allowing for a deeper understanding of the modeled world.
HDC is a promising computational model that offers high computation efficiency and noise tolerance \cite{Kanerva2023, imani2019searchd, yin2024remedy, huang2023fefet}.
Inspired by the information processing patterns of the human brain, HDC uses hyperdimensional vectors (HVs) to represent and operate data, and supports complete sets of algebraic computation.
Existing HDC models accommodate two primary representation relations, using binding-bundling form of representations (similar to disjunctive normal form) to represent objects with class-instance relation \cite{poduval2022graphd,kanerva2010} and class-class relation \cite{Hersche2023}, respectively. 
However, the development of neuro-symbolic AI demands HDC to represent more complex relations such as the class-subclass relation, where multiple objects associate  different levels of classes and subclasses \cite{Rachkovskij2013}, as shown in Fig. \ref{fig:challenge}(a).
The class-subclass relation exhibits a hierarchical structure known as ‘class-instance’,‘type-token’ or ‘is-a’ and is related to categorization or taxonomy hierarchy.

When representing the class-subclass relation, prior HDC models inevitably face the challenges for factorization, which plays a central role in the reasoning of HDC,
as shown in Fig. \ref{fig:challenge}(c).
These challenges include "superposition catastrophe" \cite{Rachkovskij2001}, where the subclass items of multiple objects  simultaneously become mixed and indistinguishable, and “the problem of 2” \cite{Gayler2006}, which induces  information loss when several identical objects are represented simultaneously.
Another challenge faced by HDC models is the inefficiency in factorization as the size of class-subclass hierarchy for representation (i.e., the number of class and subclass items) scales up.
Existing HDC models \cite{Langenegger2023,Frady2020} often rely on numerous iterations of unbinding operations and similarity measurements   to reach high factorization accuracy, resulting in   significant time costs and large number of computations.
Moreover, even when only a subset of subclasses are of interest, current HDC models still require complete factorization of the target object HVs \cite{Kleyko2023}.
These limitations hinder the neuro-symbolic application in a broader range of scenarios.

To address  above challenges, we propose FactorHD,
which consists of a symbolic encoding  that can  represent multiple objects with multiple levels of subclasses, and a factorization algorithm that can efficiently factorize encoded object representations for the selected subset of subclass items.
The main contributions of our work can be summarized as follows:

\begin{itemize}
\item 
We introduce FactorHD, a neuro-symbolic HDC model that supports the representation and factorization of  multiple objects
with class-subclass relation  hierarchy. 
Compared to  prior binding-bundling form of representations,  FactorHD employs a novel symbolic encoding method that embeds an extra  bundling clause  to connect items  and a redundant class label. 
Such bundling-binding-bundling form of representation preserves more information for 
multiple objects with multiple class and subclass levels, which can still be factorized
without causing "superposition catastrophe" and "the problem of 2".

\item 
We developed a factorization algorithm for FactorHD that offers superior computational efficiency and a reduced number of computations while maintaining high factorization accuracy. 
In contrast, prior algorithms require iterative exploration of the entire representation hierarchy. 
Additionally, our proposed algorithm allows for partial factorization of specific subclasses of interest, eliminating the need for a complete and costly factorization process.

\item 
Evaluation results demonstrate that FactorHD can factorize significantly larger representations, with faster speeds and reduced number of computations. 
The factorization time complexity of FactorHD is approximately $O(N_M)$, where $N_M$ represents the number of subclass items within each class. 
Moreover, FactorHD shows lower declining rate in factorization accuracy as the problem size grows, exhibiting high accuracy and better scalability compared to other  designs.
Integrated with the ResNet-18 neural network as the neuro part, FactorHD also shows high accuracy on training and factorizing practical datasets such as RAVEN \cite{Zhang_2019_CVPR}, Cifar-10 and Cifar-100.
\end{itemize}
\section{Background and Motivation}
\label{sec:background}
In this section, we provide an overview of HDC models, then discuss the challenges of representation and  factorization for existing HDC designs. 

\subsection{Hyperdimensional Computing Basics}
\label{sec:existing_work}
HDC was originally proposed in cognitive neuroscience for symbolic reasoning \cite{Plate1994}, based on the observation that the human brain operates on high-dimensional data \cite{Kanerva2023}. 
HDC designs are significantly powerful in reasoning and association of the abstract information, while they are weak in features extraction from complex data, e.g., image/video \cite{Poduval2021}.
Symbolic representations in HDC are based on the quasi-orthogonality characteristic of HVs, implying that
randomly generated HVs are nearly orthogonal and dissimilar to each other \cite{Kleyko2022}.

HDC measures the  similarity between HVs during inference for reasoning \cite{Kleyko2022} and factorization \cite{poduval2022graphd,Langenegger2023}.
Here we assume that $\vec{V_1}$, $\vec{V_2}$ are two randomly generated quasi-orthogonal HVs with $\vec{V_1}$,$\vec{V_2}\in\{-1,+1\}^D$, where $D$ is the dimensionality of HVs.
Common similarity metrics include cosine similarity \cite{yu2022understanding, liu2022cosime}, Hamming distance \cite{kanerva1997fully, ni2019ferroelectric, yin2023ultracompact} and dot product \cite{plate1995holographic, yin2024homogeneous}.
A similarity metric close to 0 implies that the HVs are quasi-orthogonal. 
In this design, we use dot product similarity (denoted as $sim(.,.)$)  to recognize target HVs, and the similarity is computed by $sim(\vec{V_1},\vec{V_2})=\vec{V_1} \cdot \vec{V_2} /D$.


HDC also supports complete sets of operators for computation, mainly includes bundling ($+$) that acts as memorization during HV addition, binding ($\odot$) that associates multiple HVs, unbinding that disassociate HVs, and permutation ($\rho$) that preserves the position information within a sequence \cite{ge2020classification}.

\textbf{Bundling ($+$)} of two HVs $\vec{V_1}$ and $\vec{V_2}$ is performed by component-wise addition.
The bundled HVs preserves high similarity to its component HVs, i.e., $sim(\vec{V_1}+\vec{V_2},\vec{V_1})\ \gg0$. Hence, we can identify the components by similarity measurement. 
Note that in this design, we restrict and clip the component values of bundling results of single object to the range of $\{-1,0,1\}$.
However, when bundling HVs of different objects,  we retain the results in $Z^D$ as prior designs \cite{poduval2022graphd}.

\textbf{Binding ($\odot$)} of two HVs $\vec{V_1}$ and $\vec{V_2}$ is performed by component-wise multiplication.
The resultant HV is dissimilar to any of its constituent vectors, i.e., $sim(\vec{V_1}\odot\vec{V_2},\vec{V_1})\approx0$.

\textbf{Unbinding} of HVs is achieved through the same arithmetic operation used for binding when the operation is self-inverse. 
In this case, the binding of two identical HVs results in a constant value of 1, i.e., $\vec{V_1}\odot\vec{V_1}=1$.

\subsection{Challenges  for Representation and Factorization}
\label{sec:challenge}
Existing HDC models \cite{Camposampiero2024,poduval2022graphd,Langenegger2023,NEURIPS2023minonet,kanerva2010,Hersche2023} accommodate two primary  representation relations,  using binding-bundling form to represent objects with class-instance relation \cite{kanerva2010} and class-class relation \cite{Hersche2023}. Consequently, these models can be categorized into two types based on their representation structures: the class-instance model (C-I model) and the class-class model (C-C model).
Within C-I model \cite{poduval2022graphd,NEURIPS2023minonet,kanerva2010}, objects are represented by binding class and instance (i.e., subclass) HVs.
Bundling operations are used to combine different classes.
For example, to represent a brown dog, the object is encoded as:
$object=animal\odot dog+color\odot red$,
where $animal$ and $color$ are class HVs, while $dog$ and $red$ are instance HVs.
To partially factorize the object, i.e., the $color$ of the object, the unbinding process applies:
$color\odot object = red+noise$.
As a result, the C-I model is effective in factorization and obtaining partial information. 
Within C-C model \cite{Camposampiero2024,Hersche2023,Langenegger2023}, 
binding connects different classes while bundling connects multiple objects.
For instance, the previous example combining another white cat can be represented as:
$objects=dog\odot red+cat\odot white$.
Therefore, the C-C model is capable of representing multiple objects simultaneously  \cite{Kleyko2022}.

However, when representing 
the more complex class-subclass relations, where multiple objects associate  different levels of classes and subclasses (e.g., animals-dogs-spaniels-Fido) \cite{Rachkovskij2013},  
existing HDC  models all inevitably face challenges for factorization. 
The C-I model face the challenges of “superposition catastrophe” \cite{Kleyko2022} for  multiple objects representation in the class-subclass relations, and "the problem of 2" \cite{Gayler2006} when representing several identical objects.
Even though the C-C model can avoid the above problems when representing multiple objects, they still face inefficiency in factorization as the size of representation scales up.
Given $F$ class or subclass codebooks $A_i=\{a_{i1},a_{i2},\ldots,a_{iM}\}$ with $M$ holographic bipolar item HVs ($a_{ij}\in\{-1,+1\}^D,i=1,2,\ldots,F$, $j=1,2,\ldots,M$), 
the goal is to unbind the representation product vector $H$ (e.g., $H=a_{11}\odot a_{23}\odot a_{32}$) to identify subclass items, i.e., $a_{11}$, $a_{23}$ and $a_{32}$.
Without loss of generality, when we 
factorize $a_{11}$ from $H$, 
we must identify and eliminate term $a_{23}\odot a_{32}$, necessitating exploration of all item vector combinations and similarity measurements. 
If $F=3$ and $M=256$, factorizing a specific item HV of an encoded object HV may require $M^F=16,777,216$ similarity measurements. 
As the size of representation for factorization increases, this number of repeated unbinding operation grows rapidly.
Moreover, in many scenarios where only a subset of class and subclass items are of interest, 
current C-C models still needs to factorize all composing item vectors, resulting in computation wastes.

To address the factorization challenge, various models have been proposed for the C-C model. 
The nonlinear  dynamical resonator network \cite{Frady2020} provides a neural solution to the factorization problem by iteratively searching for the
$F$ terms across the set of possible solutions, assuming that each codebook contains a finite set of $M$ code vectors. 
Another approach \cite{Langenegger2023} designed a novel in-memory stochastic factorizer (IMC factorizer) to handle larger problem sizes  (i.e., size of representation hierarchy) with high accuracy. 
At $D=256$, $F=3$ and $M=256$, the resonator network \cite{Frady2020} becomes incapable of factorizing the representation vectors, while the IMC factorizer \cite{Langenegger2023} achieves an accuracy of $99.71\%$ with an average  3312 iterations. 
Nevertheless, 
C-C models still exhibit relatively low computing efficiency when factorizing larger problems.

\section{FactorHD for Multi-Object  Multi-Class Representation and Factorization}
\label{sec:schematic}

 
In this section, we introduce FactorHD, a novel neuro-symbolic  HDC model that 
provides enhanced  symbolic representation and factorization capability with complex hierarchy. 
We discuss the encoding and factorization method for 3 different symbolic representations of FactorHD.
We finally investigate the selection of optimal $TH$ values for factorizing representations with multiple objects.

\subsection{FactorHD Encoding for Representation}

Here we introduce the proposed symbolic encoding method for FactorHD representation with class-subclass relation hierarchy.
Fig. \ref{fig:challenge}(a) shows an example of the proposed class-subclass relation hierarchy with two subclass levels represented in bundling-binding-bundling form.
In practical scenarios, 
fixed labels are assigned to each class to distinguish different classes.
We use $LABELi$ to denote the HVs of the top  class $a_i$, as the top  level class is often used as the label in the traditional class-instance  classification.
The class-subclass relation of different levels in a class is known in advance, such as “dog is animal”, and notably, our focus is on identifying whether an object contains an item of a specific class or subclass, rather than determining hierarchical levels.

To encode objects in FactorHD,  different subclass levels belonging to the same class can be considered equally and combined together using the bundling operation, and different classes are linked using the binding operation. 
We then use the bundling operation to connect different  objects just as a class-class structure,  forming a bundling-binding-bundling structure. 
An example of FactorHD encoding is 
$H=(LABEL1+a_{11}+a_{112})\odot(LABEL2+a_{22}+a_{221})\odot\ldots$,
which includes redundant class labels $LABEL1$, $LABEL2$, and subclass items  $a_{11}$, $a_{112}$,  $a_{22}$ and $a_{221}$, etc. 
Among them, $a_{11}$ belongs to class $a_1$, $a_{112}$ belongs to $a_{11}$, and $a_{22}$ belongs to class $a_2$, $a_{221}$ belongs to $a_{22}$. 
Note that even if the items of a specific class is not associated with an object,  FactorHD  still reserves its class label and bundles the label with a global HV representing 'NULL'.
With such encoding approach, FactorHD eliminates the need for foreknowledge of  specific classes associated in the object,
while C-C models require exhaustive factorization if contained class is unknown
\cite{Langenegger2023,Frady2020}.
Below we elaborate on how our proposed FactorHD model factorize different encoded object representations with class-subclass relation.

\begin{figure}
     \centering
     \includegraphics[width=1\linewidth]{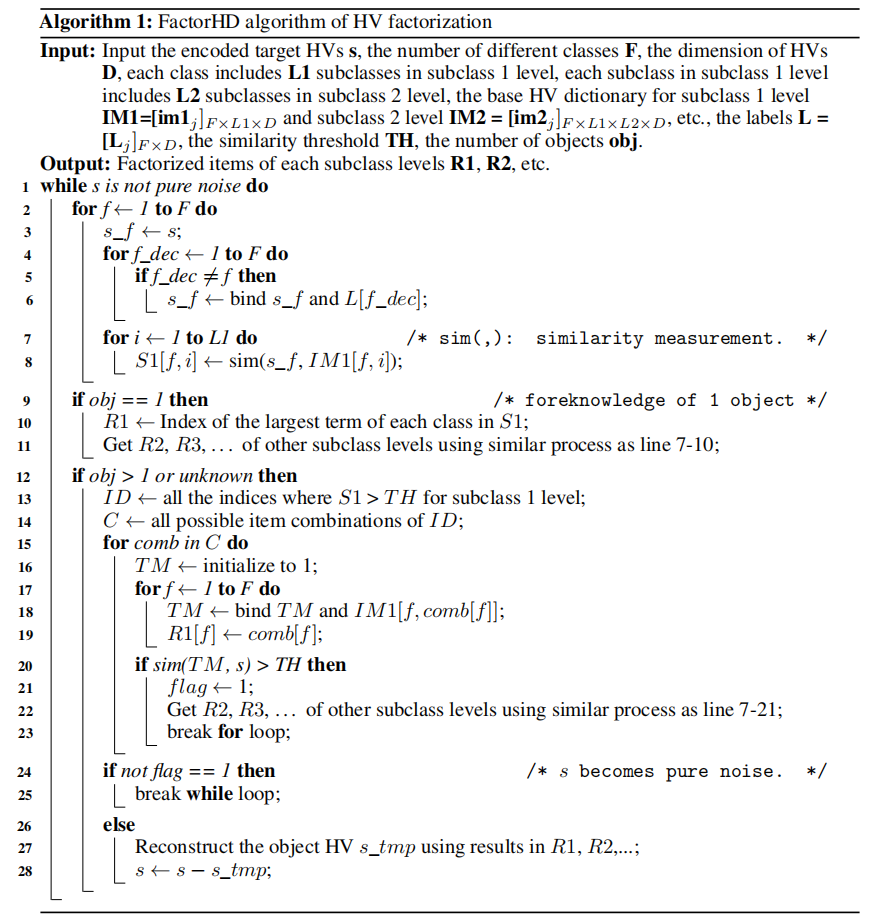}
 \label{algorithm1}
 \vspace{-4ex}
 \end{figure}

\subsection{FactorHD Factorization Algorithm}\label{sec:basic decoding}

Algorithm. 1 summarizes the proposed factorization algorithm of FactorHD.
The loop starts from the original target HV.
For each selected class of the object, we first unbind all other unselected class labels.
Then, we compute the similarity between the  unbound HV and all subclass items of the selected class.
When  the target HV only represents single object,
we simply select the subclass item with the largest similarity to the unbound HV as the factorized subclass item of the selected class, 
and repeat the similarity measurement and subclass item selection process for all subclass levels. 
On the other hand, when the target HV represents multiple objects, or the number of objects is unknown,   
we need to select all the subclass items at the same subclass level with a similarity  larger than a predefined "threshold similarity" ($TH$) to  their respective unbound class HVs, as these subclass items are possibly associated with the multiple objects.
These subclass items are then selectively  bound, such that each of the resultant product HVs  contain exactly one subclass item of every subclass. 
If one of the combinations is similar to the target HV, i.e., the similarity is larger than $TH$, it indicates that the target HV contains an object possessing the subclass items of this specific combination at this subclass level.
We then repeat the similarity measurement and subclass item selection process for lower subclass levels.
Only when the subclass items at all subclass levels are determined, we infer that the corresponding  object possessing all these subclass items exists in the target HV. 
The HV of the object will then be reconstructed using the factorized class labels and subclass items, and excluded from the target HV. 
The new target HV will move to the next loop  until no more objects can be factorized from the target HV.

To further illustrate the algorithm, we  discuss the factorization  for 3 different target class-subclass representations,  from the common case to  more complex cases.

\begin{figure}
     \centering
     \includegraphics[width=1\linewidth]{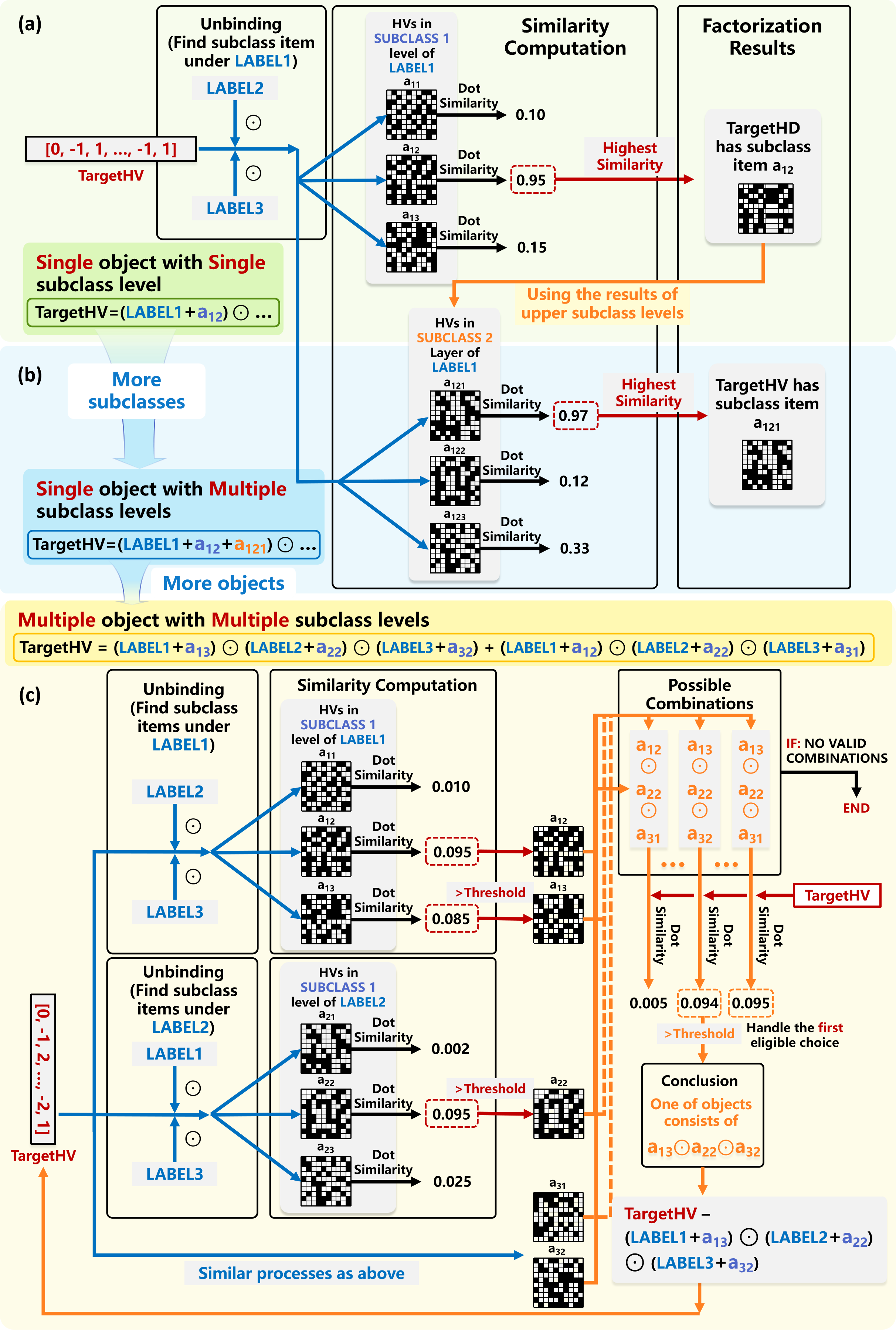}
     \vspace{-2em}
     \caption{(a) Illustration of FactorHD factorizing the common Rep 1 case. 
     (b) Factorizing  single object with multiple subclass levels (Rep 2). (c) Factorizing the  multiple objects with multiple subclass levels (Rep 3). 
     }
 \label{fig:decoding}
 \vspace{-3ex}
 \end{figure}

\textbf{Representation 1 (Rep 1)}: Fig. \ref{fig:decoding}(a) shows the factorization process for single object with single subclass level.
The unconsidered class labels can be eliminated by binding the same label:
$(LABEL2+a_{21})\odot LABEL2\approx1$.
In this way, 
we factorize the information associated with the selected class:
\begin{equation}
\label{eq1}
TargetHV\odot LABEL2\odot\ldots\approx(LABEL1+a_{12})
\end{equation}
Eq. \ref{eq1} shows the unbinding result  only contains the bundling clause with the selected class $LABEL1$.
By computing the similarity between the unbinding result with all the subclass item HVs 
of class $LABEL1$, we can easily identify the subclass item with the highest similarity as the factorized one.  



\textbf{Representation 2 (Rep 2)}: Fig. \ref{fig:decoding}(b) illustrates the  factorization process for single object with multiple (e.g., 2) subclass levels.
On the basis of Rep. 1, after factorizing out the item at subclass 1 level, i.e., $a_{12}$, 
we compute similarity specifically with the item HVs under $a_{12}$,
and  select the item HV with the highest similarity again, i.e., $a_{121}$, as the factorized item at subclass 2 level. 
Such process can continue to subsequent subclass levels, which
minimizes the number of similarity computations required to factorize all subclass items of the target object by searching from higher to lower levels.

\textbf{Representation 3 (Rep 3)}:
In the most complex scenario, we handle multiple objects with multiple subclass levels. Current methods compare the target HV with all subclass combinations, which is computationally intensive. 
When factorizing target HV with multiple objects expressed in Fig. \ref{fig:decoding}(c) using FactorHD, simply binding all unselected class labels with $TargetHV$ is not sufficient, as multiple subclass items under the same class  remain. 
For instance, binding $TargetHV$ with $LABEL2$ and $LABEL3$ results in multiple subclass items of class $LABEL1$, i.e., $a_{13}$ and $a_{12}$, and selecting one subclass item each time may cause "superposition catastrophe". 
Therefore, FactorHD only considers items with a similarity higher than $TH$. 
After unbinding classes and selecting subclass items, though we cannot determine specific item combinations,
we can determine that the class  $LABEL1$ contains $a_{12}$ and $a_{13}$, the class  $LABEL2$ contains $a_{22}$, and the class  $LABEL3$ contains $a_{31}$ and $a_{32}$. 
Theoretically, only 4 subclass item combinations need to be considered (e.g., $a_{12}\odot a_{22}\odot a_{31}$).
If any combination exceeds $TH$, we conclude that $TargetHV$ contains that specific combination.
Further processes have been detailed in Algorithm. 1.
FactorHD significantly reduces the number of combinations needed for similarity computation, 
addresses the "superposition catastrophe" \cite{Rachkovskij2001} and “the problem of 2” \cite{Gayler2006}, 
and adapts to any number of objects.

\begin{figure}[!t]
     \centering
     \includegraphics[width=1\linewidth]{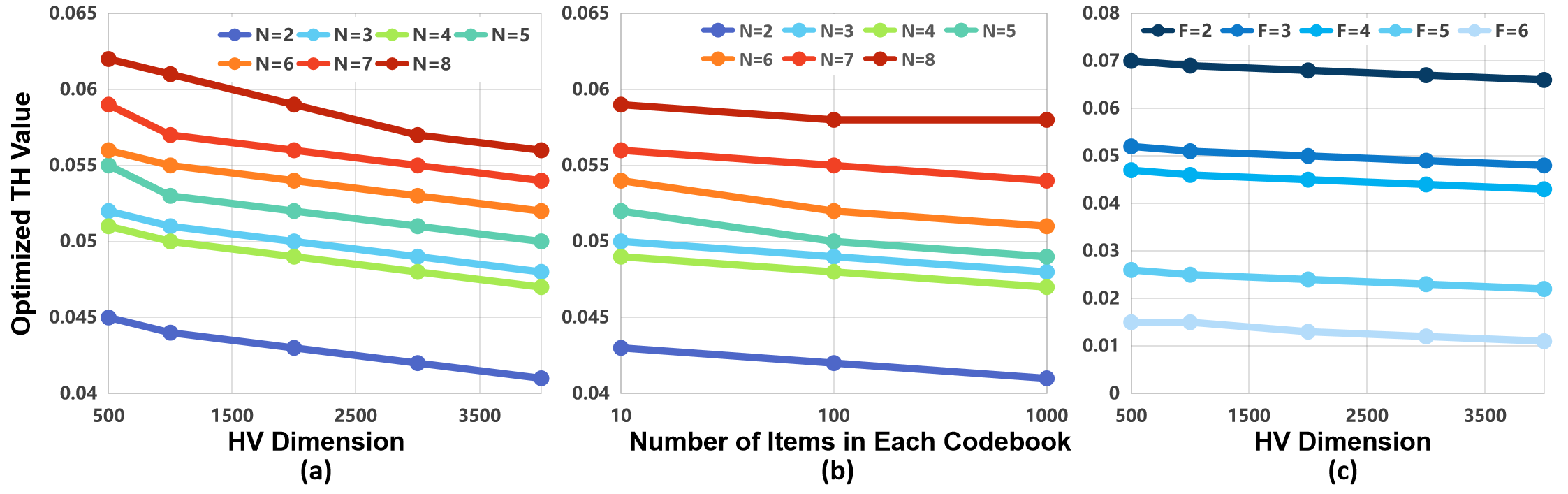}
     \vspace{-2em}
     \caption{The optimal value of $TH$ for factorizing representations of multiple objects with a single subclass level varies with (a)  HV dimension $D$ and the number of objects $N$ given $M=10$ and $F=4$, (b)  the number of items in each codebook $M$ given $D=2000$ and $F=4$, and (c)  the number of factors $F$ in the composite HVs given $N=3$ and $M=10$.
     }
 \label{fig:THvalue}
 \vspace{-3ex}
 \end{figure}

Fig. \ref{fig:THvalue} illustrates the method for selecting the optimal $TH$ value $TH^*$ across specific factorization problems. 
$TH^*$  is defined as the value such that FactorHD achieves the highest factorization accuracy on Rep 3.
$TH^*$ is determined by HV dimension, representation structure, and factorization problem complexity, increases with the number of objects $N$ and decreases with the number of factors $F$. Moreover, $TH^*$ shows a roughly linear relationship with HV dimension $D$ and the logarithmic value of $M$. 
Therefore, given $N$ and $F$, $TH^*$ can be approximately fitted and predicted. And values near $TH^*$, though not the optimal, also yield high factorization accuracy. 
Thus, a rough value selection can be fit as follows:
\begin{equation}
    TH^{*}=0.001(104+2N-15F-0.001D-log(M))
\end{equation}
In contrast, the performance of state-of-the-art (e.g., the IMC Factorizer) heavily depends on the selection of specific activated values and their thresholds \cite{Langenegger2023}, even for factorizing Rep 1, and these values are completely irregular.


\section{Performance Evaluations}
In this section, we validate and evaluate the performance of FactorHD.
We first compare it with other HDC models on the common Rep 1 problem.
Then, we evaluate the factorization accuracy of FactorHD on Rep 2 and Rep 3, as well as on practical datasets such as RAVEN, Cifar-10, and Cifar-100.

\label{sec:eval}
\vspace{-1ex}
\subsection{Experimental Setup}
\label{sec:decoding1}
Algorithmic experiments are implemented on the GeForce RTX 2080 GPU platform. For the basic experiments on Rep 1-3, we perform 1024 factorization trials respectively, with batch size of 512.
We first evaluate the performance of FactorHD on Rep 1, comparing with C-C models on factorization accuracy and time cost.
The HV dimensions used here are consistent with those in \cite{Frady2020,Langenegger2023,Kent2020}, namely $D=1500$ for $F=3$ and $D=2000$ for $F=4$. 
FactorHD operates in $\{-1,0,1\}^D$ space, using 2 bits per dimension,
thus  $D$ of FactorHD reduces by half to match the storage space of other models to ensure fair comparisons. 
The number of code vectors in each class codebook $M$ is varied, determining the total problem size as $M^F$.
Then when comparing to C-I models \cite{Camposampiero2024}, we set $D=256$ for $F=3$ and $D=512$ for $F=4$ in the factorization of C-I models. We similarly reduce $D$ for FactorHD by half.

Given that prior HDC designs fail to accommodate complex representations, 
we simply evaluate FactorHD on Rep 2 and 3. 
The testing representations include one or two objects, each with two subclass levels. The top-level classes consist of 256 subclasses, each having 10 sub-subclasses.

Finally, we present the training and factorization performance of FactorHD with the ResNet-18 network as the neuro part for feature extraction, using the RAVEN \cite{Zhang_2019_CVPR}, Cifar-10, and Cifar-100 datasets. 
The RAVEN dataset includes objects characterized by attributes like type, size, color, and position.
The first codebook represents the position attribute, the second represents color, and the third combines size and type attributes, resulting in 30 size-type combinations.
Image patterns in RAVEN dataset consists of 1-9 objects.
Cifar-10 images are encoded by binding the image label with a dummy label. 
Cifar-100 datasets naturally have two class levels (coarse and fine labels), the two levels of labels are bound for encoding, and later
factoring out either the coarse or fine information for specific images.
During training, images containing both coarse and fine aspect information are fed into neural network to extract features, which are then encoded into HVs.
During inference, HVs are factorized to reveal the specific composition of the images.


\begin{figure}[!t]
     \centering
     \includegraphics[width=1\linewidth]{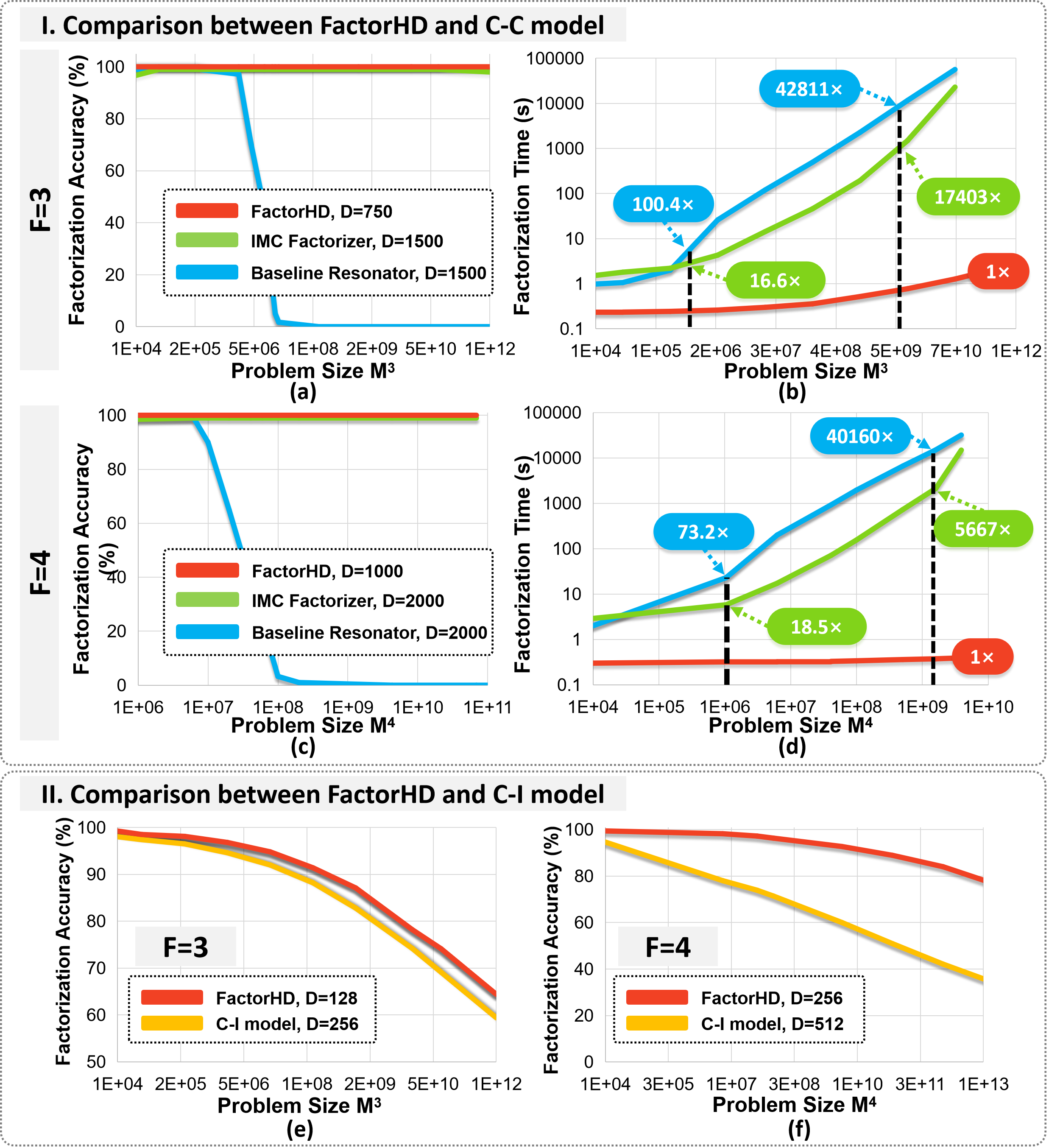}
     \vspace{-2em}
     \caption{I. The factorization performance comparison between FactorHD and C-C models using baseline resonator network, IMC factorizer with varying HV dimensionality:
     When $F=3$, (a)
     the 
     factorization accuracy and (b) factorization time with scaling problem size $M^3$. 
     When $F=4$,  (c) the factorization accuracy and (d) factorization time with scaling problem size $M^4$. 
     II. The factorization performance comparison between FactorHD and C-I models when (e) $F=3$ and (f) $F=4$ with varying HV dimensionality.
     }
 \label{fig:f34}
 \vspace{-1.5em}
 \end{figure}

\vspace{-1ex}
\subsection{Experimental Results}
\vspace{-0.5ex}

\textbf{Rep 1.} Fig. \ref{fig:f34}(a-d) show the  factorization accuracy and factorization time of FactorHD and C-C models using different factorization methods with $F=3$ and $F=4$ , respectively.
The results suggest that FactorHD maintains a high factorization accuracy above $99\%$ even at lower HV dimensions, maintaining stability as problem size increases.
On the contrary, the C-C model using resonator network fails when the problem size reaches $10^6$, and
the IMC factorizer drops from the accuracy of $99\%$ at problem size of $10^{12}$.

Moreover, FactorHD exhibits exceptional factorization efficiency. 
By employing a novel symbolic encoding method that embeds an extra HDC bundling clause and factorization algorithms,
the factorization process requires much less iterations, resulting in significantly reduced time cost and making the factorization process well-suited for parallel computing. 
Consequently, as the problem size grows, the factorization time increase of FactorHD is negligible compared to approaches factorizing C-C models.
The time complexity of FactorHD is approximately $O(N_M)$, where $N_M$ represents the maximum number of subclass items within each class, while 
the IMC factorizer \cite{Langenegger2023} and resonator network \cite{Frady2020} have a time complexity larger than $O(N_M^2)$.
Thus, the speedup of FactorHD over existing models on factorization grows with problem size, achieving a minimum speedup of $18.5\times$ at $10^6$ problem size and reaching $5667\times$ at $10^9$ problem size.

Fig. \ref{fig:f34}(e) and (f) further demonstrate the  factorization accuracy of FactorHD and the baseline C-I model with $F=3$ and $F=4$ , respectively.
The results show that FactorHD achieves higher factorization accuracy across various problem sizes compared to the simpler C-I model, demonstrating its ability to maintain high accuracy while representing more complex structures.
The factorization times for both models are comparable across different problem sizes due to the  similar algorithms of label elimination during factorization.

\begin{figure}[!t]
     \centering
     \includegraphics[width=1\linewidth]{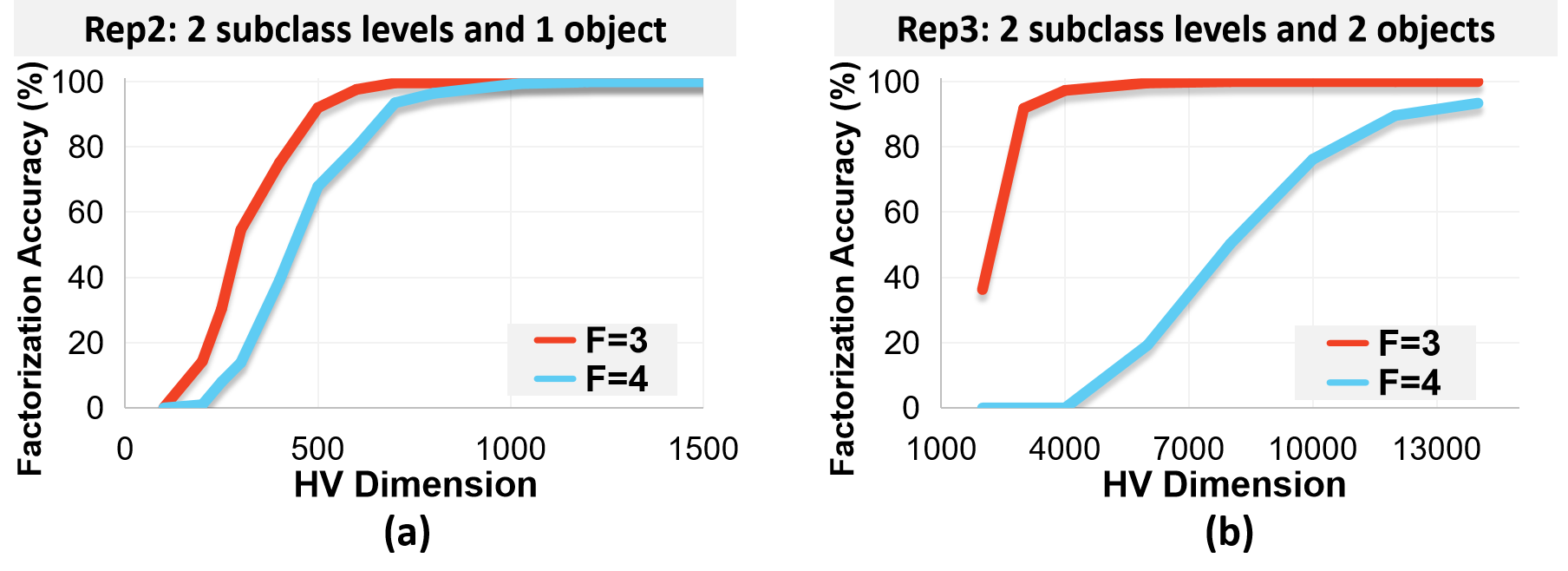}
     \vspace{-2.5em}
     \caption{The performance of FactorHD on factorizing (a) Rep 2 and (b) Rep 3 with varying HV dimensionality. 
     }
 \label{fig:rep23}
 \vspace{-1em}
 \end{figure}

\textbf{Rep 2\&3.} 
Fig. \ref{fig:rep23}(a) shows the FactorHD performances on factorizing Rep 2. Compared to Rep 1, factorizing representations with multiple subclass levels requires higher HV dimensions due to larger problem sizes and more complex structures. Nevertheless, factorization accuracy reaches $100\%$ at $D=1000$.
Fig. \ref{fig:rep23}(b) shows the FactorHD performances on factorizing Rep 3. In this case, the algorithm does not require prior knowledge of the actual number of objects, necessitating higher HV dimensions to achieve high factorization accuracy.

\begin{table}[t!]
\centering
\caption{Factorization accuracy on RAVEN test set}
 \vspace{-3mm}
\label{tab:raven_result}
\centering
    \includegraphics[width=1\linewidth]{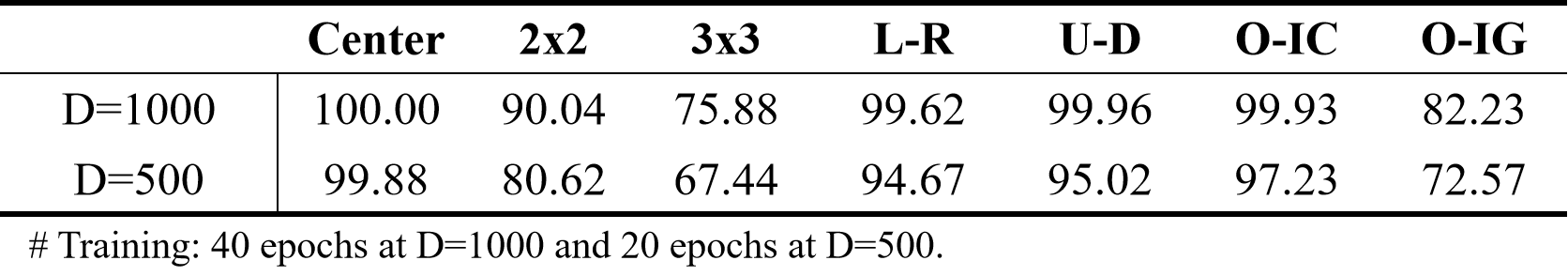}
    \vspace{-3em}
\end{table} 

\begin{table}[t!]
\centering
\caption{Factorization accuracy on Cifar-10 and Cifar-100 test set}
 \vspace{-3mm}
\label{tab:cifar}
\centering
    \includegraphics[width=1\linewidth]{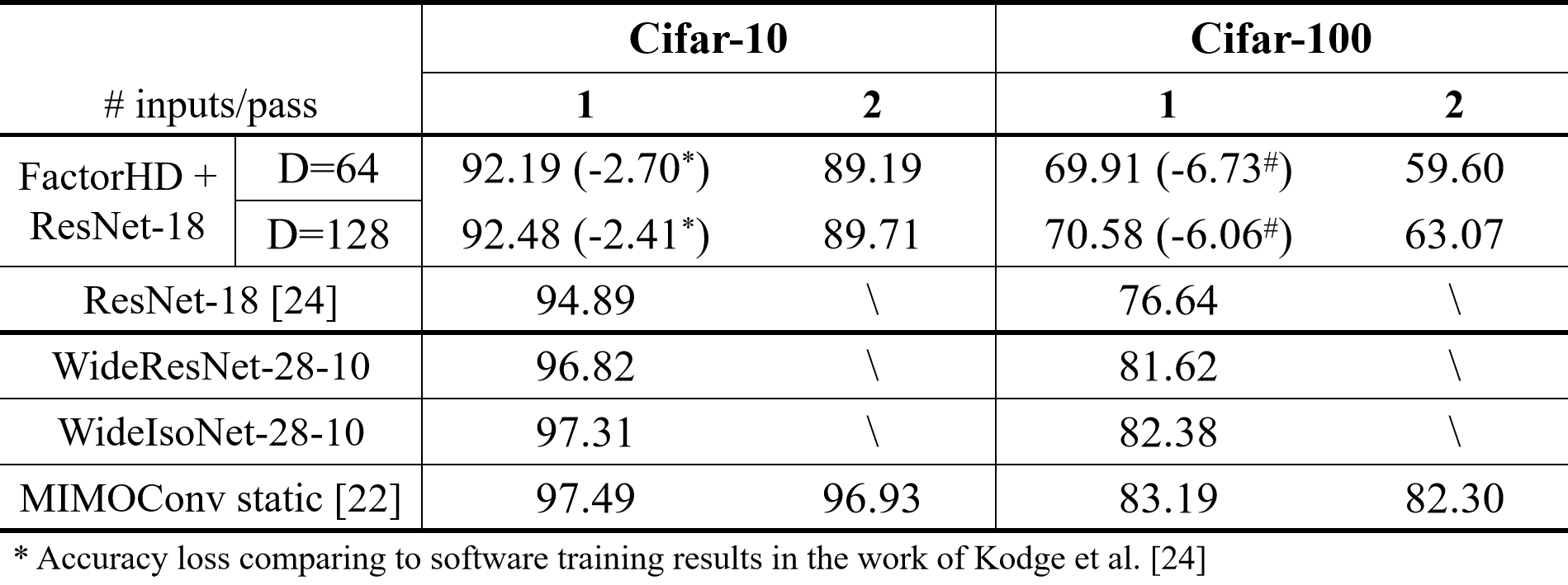}
    \vspace{-3.5em}
\end{table} 

\textbf{Inference accuracy.}
We evaluate FactorHD's accuracy on the RAVEN \cite{Zhang_2019_CVPR}, Cifar-10 and Cifar-100 datasets.  
Table \ref{tab:raven_result} shows the accuracy of  factorizing RAVEN test sets. 
Training and factorization with $D=1000$ achieve over $90\%$ accuracy for most RAVEN patterns.
With reduced HV dimensionality, decent accuracies are still maintained.
Table \ref{tab:cifar} further shows the factorization accuracy of FactorHD integrated with ResNet-18 on the Cifar-10 and Cifar-100 datasets.
Compared to the classification accuracy using ResNet-18 neural network \cite{kodge2024deep} and other emerging neural networks (e.g., WideResNet-28, WideIsoNet-28) in traditional neural classification tasks, the accuracy loss of FactorHD model integrated with ResNet-18 is less than $3\%$ for Cifar-10. 
The accuracy loss can be further minimized with an increase in HV dimension.
We also adjust the number of bundled image inputs during training to assess the ability to factorize  superposition \cite{NEURIPS2023minonet}.
FactorHD still gets high factorization accuracy with image inputs in superposition, particularly in the simpler Cifar-10 dataset, thus improving the training efficiency with reduced training epochs.
Compared to MIMOConv \cite{NEURIPS2023minonet} that specifically designed for bundled images training, the testing results demonstrate the stability of our FactorHD, maintaining relatively high accuracy even integrated with simpler neural networks, i.e., ResNet-18.

\vspace{-1ex}
\section{Conclusion}
\label{sec:conclusion}
\vspace{-1ex}
In this article, we propose FactorHD, a novel neuro-symbolic model that overcome the challenges of  HDC   representation and  factorization  for multiple objects with multiple subclass levels.
FactorHD employs a novel symbolic encoding method,
coupled with an efficient factorization algorithm that efficiently identifies items of interest without exhaustive similarity measurements.
Experimental evaluations demonstrate that FactorHD not only improves speed and accuracy for the common representation but also effectively handles complex representations involving multiple objects and subclass levels.
Compared to existing HDC models on factorizing the common representation, i.e., single object with single subclass level, FactorHD achieves approximately $18.5\times$ speedup at the problem size of $10^6$, and   $5667\times$ speedup at the problem size of $10^9$, while maintaining  $99\%$ accuracy.
Furthermore, FactorHD integrated with ResNet-18 neural network achieves $92.48\%$ factorization accuracy on the Cifar-10 dataset.

\vspace{-1ex}
\section*{Acknowledgment}
This work was partially supported by NSFC (92164203) and SGC Cooperation Project (Grant No. M-0612).
\small
{
\bibliographystyle{ieeetr}
\bibliography{bib}
}


\end{document}